\documentclass[10pt,
aps, 
reprint, 
twocolumn,
superscriptaddress, 
notitlepage,
amsmath,amssymb 
]{revtex4-2}
\usepackage[utf8]{inputenc}
\usepackage{natbib}
\usepackage{graphicx}
\usepackage{url}
\usepackage[normalem]{ulem}

\usepackage[table]{xcolor}
\usepackage{tikz}
\usepackage[ngerman,english]{babel}
\usepackage{mdframed}
\usepackage{lineno}

\newcommand{\diff}{\,\textnormal{d}}

\newcommand{\lmax}{\,\ell_\mathrm{max}}
\newcommand{\ttil}{\,\tilde{t}}
\newcommand{\vp}{\,v_\mathrm{p}}
\newcommand{\vv}{\,v_\mathrm{b}}
\newcommand{\ltil}{\,\tilde{L}}
\newcommand{\lcp}{\,\langle \ell\rangle_\textnormal{cp}}
\newcommand{\lav}{\,\langle \ell\rangle}
\newcommand{\lavr}{\,\langle \ell(r)\rangle}
\newcommand{\lavtr}{\,\langle \ell(\tilde{r})\rangle}
\newcommand{\oav}{\,\langle o\rangle}
\newcommand{\omd}{\,\omega_\mathrm{d}}
\newcommand{\omi}{\,\omega_\mathrm{i}}
\newcommand{\omn}{\,\omega_\mathrm{r}}
\newcommand{\ome}{\,\omega_\mathrm{s}}
\newcommand{\tr}{\,\tilde{r}}

\newcommand{\tcp}{\,\tilde{t}_\mathrm{walk}^\mathrm{max}}

\begin{document}
	\title{Dynamic stop pooling for flexible and sustainable ride sharing}
	
	\author{Charlotte Lotze}
	\affiliation{Chair for Network Dynamics, Institute for Theoretical Physics and Center for Advancing Electronics Dresden (cfaed), Technical University of Dresden, 01069 Dresden}
	
	\author{Philip Marszal}
	\affiliation{Chair for Network Dynamics, Institute for Theoretical Physics and Center for Advancing Electronics Dresden (cfaed), Technical University of Dresden, 01069 Dresden}
	
	\author{Malte Schröder}
	\affiliation{Chair for Network Dynamics, Institute for Theoretical Physics and Center for Advancing Electronics Dresden (cfaed), Technical University of Dresden, 01069 Dresden}
	
	\author{Marc Timme}
	\affiliation{Chair for Network Dynamics, Institute for Theoretical Physics and Center for Advancing Electronics Dresden (cfaed), Technical University of Dresden, 01069 Dresden}
	\affiliation{Lakeside Labs, 9020 Klagenfurt am W{\"o}rthersee, Austria}
	\email{marc.timme@tu-dresden.de}

	\begin{abstract}
		Ride sharing -- the bundling of simultaneous trips of several people in one vehicle -- may help to reduce the carbon footprint of human mobility. However, the complex collective dynamics pose a challenge when predicting the efficiency and sustainability of ride-sharing systems. Standard door-to-door ride sharing services trade reduced route length for increased user travel times and come with the burden of many stops and detours to pick up individual users. Requiring some users to walk to nearby shared stops reduces detours, but could become inefficient if spatio-temporal demand patterns do not well fit the stop locations. Here, we present a simple model of dynamic stop pooling with flexible stop positions. We analyze the performance of ride sharing services with and without stop pooling by numerically and analytically evaluating the steady state dynamics of the vehicles and requests of the ride sharing service. Dynamic stop pooling does a-priori not save route length, but occupancy. Intriguingly, it also reduces the travel time, although users walk parts of their trip. 
		Together, these insights explain how dynamic stop pooling may break the trade-off between route lengths and travel time in door-to-door ride sharing, thus enabling higher sustainability and service quality.   
	\end{abstract}
	\maketitle
	
	\section{Introduction}
	Emergent collective dynamics make it difficult to understand and predict the behavior of complex systems.
	\cite{mahan2013many, JarosawKwapien.2012, holovatch2017complex, newman2018networks,  strogatz2001exploring, Strogatz.2005, makse1995modelling,de2013mathematical, wang2017unification}. For instance in mobility systems, many different agents with various aims interact, which makes it hard to quantify key indicators like the efficiency. Methods from statistical physics, like network theory \cite{santi2014_shareabilityNetworks}, scaling analysis \cite{tachet2017scaling, Molkenthin.2020}, or mean-field theory \cite{herminghaus2019mean} can help to overcome this challenge. 
	
	In human mobility in particular, understanding the efficiency of different services is crucial to enable a shift towards more sustainable mobility. 
	Individual motorized mobility is highly inefficient with only about $1.3$ passengers per car on average \cite{Lenz.2010}. Making human mobility more sustainable requires a reduction of total route length driven and simultaneously fewer numbers of vehicles. Arguably, the most influential factor towards achieving this goal is a substantial increase of the average number of passengers per vehicle.
	
\begin{figure*}
		\includegraphics[width=\textwidth]{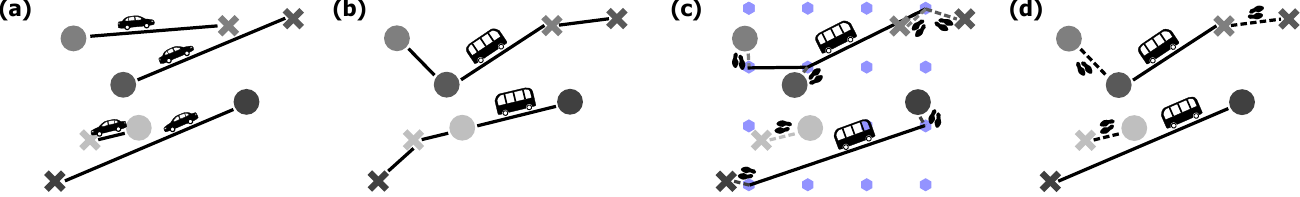}
		\vspace{-0.5cm}
		\caption{\textbf{Route length and travel time depend on mode of transport.}
			\textbf{(a)} individual mobility is the fastest mode of transport but requires longest total route length. \textbf{(b)} Standard ride sharing serves users from door to door. Ride sharing reduces total route length by combining trips and requires fewer (but larger) buses. Users may become slower due to detours and between stops. \textbf{(c)} Static stop pooling with fixed stop positions (purple hexagons) reduces the number of stops but requires users to always walk part of their trip which might increase travel time even further. \textbf{(d)} Dynamic stop pooling with flexible stop positions combines efficient public transportation and adaptive ride sharing and might thus save both total route length and user travel time despite users might walk a part of their trip.
		}
		\label{fig:sketch}    
		\vspace{-0.5cm}
	\end{figure*}

	\textit{Ride sharing} (also called ride pooling) \cite{alonso2017demand, santi2014_shareabilityNetworks, Molkenthin.2020, Ruch.2020, Agatz.2012} constitutes a promising tool to bundle multiple user trips in a single vehicle - for instance micro- or minibuses with typically 4 to 24 seats \cite{cervero2000}. While each individual user incurs a small detour on their trip, the ride sharing buses serve the users with a significantly shorter route length than in individual mobility  where each user drives in their own car \cite{Ruch.2020} (Fig.~\ref{fig:sketch}a~and~b). 
	
	However, many small detours to pickup users individually in door-to-door ride sharing services increase both the total route length (reduced sustainability) as well as user travel times (reduced service quality). \textit{Stop pooling} offers the possibility to reduce these detours: 
	if users walk a short distance to nearby stops (as in public transportation), the buses stop less often and save some door-to-door detours \cite{Yao.2020, Lin2016, Ma.2013, Stiglic2015, Fielbaum.2021,Mounesan.25.04.2021}.
	
	Fixed positions of stops as in line bus services enable a simple implementation of stop pooling. Each user walks to and from the closest stop, reducing the number of possible combinations of trips and thus the computational effort of the algorithm that bundles the trips.
	Such a \textit{static} implementation of stop pooling with fixed, prescribed stops (Fig.~\ref{fig:sketch}c) reduces the relative route length but typically increases the user travel time  \cite{Yao.2020, Stiglic2015, Lin2016, Ma.2013}.  To overcome this challenge, we here propose \textit{dynamic} stop pooling, where both bus route and user stops are adapted to current demand (Fig.~\ref{fig:sketch}d). Two recent algorithmic 
	models on dynamic stop pooling \cite{Fielbaum.2021, Mounesan.25.04.2021} suggest the possibility for both shorter total route length and simultaneously shorter travel time but do not analyse the mechanisms underlying this observation.
	
	Most studies of ride sharing services focus on operational aspects, including user behavior and economics \cite{Storch.2021, Ke.2020, WANG2019122, Ruch.2020} or algorithmic optimization \cite{alonso2017demand, wallar2018vehicle,Lobel2020}, especially in contrast to individual mobility. 
	Recent studies have begun to develop an understanding of the collective dynamics of ride sharing fleets from a complex systems perspective, revealing how these dynamics impact the efficiency of ride sharing across settings \cite{santi2014_shareabilityNetworks, tachet2017scaling, herminghaus2019mean, Molkenthin.2020}. However, an analysis of the collective dynamics induced by dynamic stop pooling and their effect on the ride sharing service quality is still missing. 
	
	In this article we present, first, a simple multi-agent model for ride sharing that captures the trade-off between route length and travel time in door-to-door ride sharing; second, include dynamic stop pooling and show how it may decrease the travel time by reducing detours between stops, and third, demonstrate how this enables dynamic stop pooling to break the ride sharing trade-off. We conclude that dynamic stop pooling may improve both route length and travel time simultaneously by adjusting the maximal walk distance of the users and the number of buses. 
	Dynamic stop pooling could thus allow to establish a fast, flexible \textit{and} sustainable ride sharing service.

	\section{Model}
	\label{sec:model}
	\subsection{Ride Sharing}
	
	The collective dynamics of ride sharing is determined by the interaction of user requests and the buses serving them.
	Let us consider the following simple model for ride sharing: users request a service to transport them from their origin to their destination as soon as possible; 
	the service provider operates a fleet of $B$ buses to serve these users; when a request is posed, it is assigned to a bus according to an assignment algorithm (see Sec.~\ref{sec:setting}). That is, origin and destination are inserted at appropriate positions into the current route of the bus as pickup and drop-off stops. 
	In the model the order of the scheduled stops once assigned does not swap, even if later requests are inserted into the bus route. 
	Over time, the buses drive with velocity $\vv$ and visit all scheduled stops one after each other (Fig.~\ref{fig:stop_pooling_vs_Ba}, left panel).

	\subsection{Dynamic Stop Pooling}
	
	\begin{figure*}[t]
		\includegraphics[width=\textwidth]{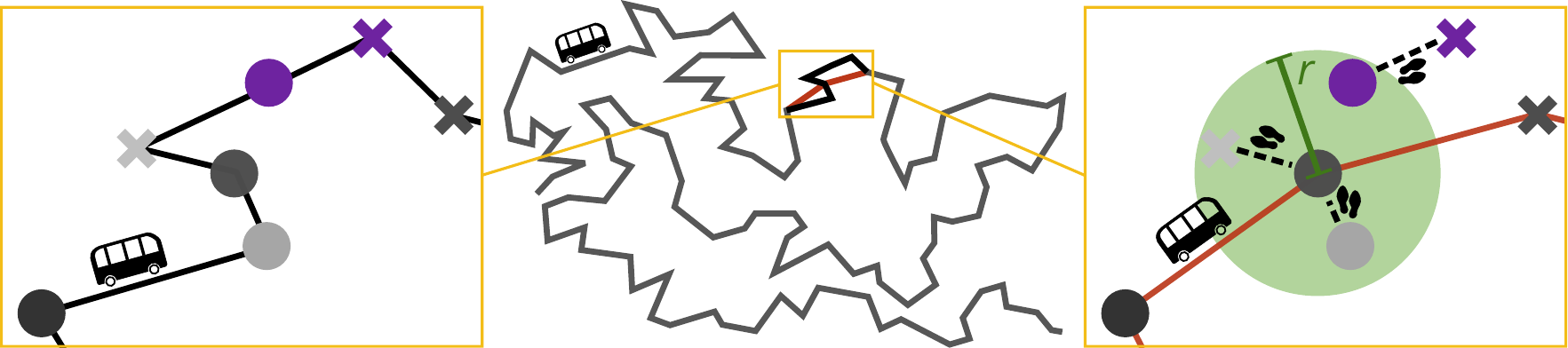}
		\caption{\textbf{Dynamic stop pooling avoids door-to-door detours by combining close-by stops.}
			In door-to-door ride sharing (Left), the bus drives to each stop resulting in 
			an overall route with many small detours (black line). Dynamic stop pooling may avoid these detours by combining close-by stops, yet keeps the overall route structure similar (compare red line). In detail, some users walk to a nearby stop closer than pool radius $r$ (Right). Their stops are \textit{served indirectly}. If origin and destination are closer than ${2\,r}$, users walk completely and do effectively not use the service. Their stops are \textit{rejected}. Ultimately, the bus serves only the remaining stops \textit{directly}.
		}
		\label{fig:stop_pooling_vs_Ba}    
	\end{figure*}
	
	With dynamic stop pooling, users may have to walk a short distance at their origin and destination - at most \textit{pool radius} $r$ per stop. Users walk from their origin to a close stop, which has to be already planed, if they reach it before the bus; similarly, they walk from a close stop to their destination. Thus, the buses can serve multiple users at one stop and save stops and related door-to-door detours.
	
	Stops are either served directly or indirectly, or rejected (not served). 
	If \textit{served directly}, the user is picked up or dropped off directly at their desired stop; if \textit{served indirectly}, the user walks to or from a close directly served stop.
	To avoid users to walk further than their requested trip length and to waste time (both their own and that of the service fleet), people with requested trip length~$\ell<2\,r$~are~\textit{rejected} and walk completely. 
	In contrast to directly and indirectly served stops, their stops are not served by the buses. 
	
	Figure \ref{fig:stop_pooling_vs_Ba} illustrates the difference between dynamic stop pooling and door-to-door ride sharing as well as the resulting stop types. These stop types yield three distinct user types: If both origin and destination are served directly, users \textit{do not walk}; if one or two stops are served indirectly, users \textit{walk partially}; otherwise the request is rejected and users \textit{walk completely}.

	\subsection{Setting}
	\label{sec:setting}
	We model the dynamics of the ride sharing service by Monte Carlo simulations \cite{mooney1997monte}. 
	The requests follow a Poisson process \cite{fischer1993markov} with mean field request rate $\lambda$ where the position of the origin is distributed uniformly in a unit square with periodic boundaries. Destinations are distributed uniformly in a disk around the origin with maximal trip length $\lmax=1/2$ such that diagonal trips are not more probable than others. 
	The trip length ($\textnormal{tl}$) of all users is thus distributed according to
	
	\noindent \begin{align}
		\rho_{\textnormal{tl}}(\ell) = \frac{2}{\lmax^2} \ell \,, \qquad \int\limits_0^{\lmax} \rho_{\textnormal{tl}}(\ell) \diff \ell=1\,
		\label{eq:trip_length}
	\end{align}
	with an average trip length $\langle \ell\rangle = 1/3$ (see Supplementary Material~A, Eq.~(S2)). 
	
	We introduce stop pooling with the same pool radius for all users, independent of their trip length. To avoid that users walk further than their trip length, we reject users with $\ell<2\,r$, where the factor $2$ captures the fact that users may walk a distance $r$ from their origin as well as to their destination. 
	We rescale the pool radius as
	
	\noindent \begin{align}
		\tr=\dfrac{2\,r}{\lmax} \,,
		\label{eq:relativer}
	\end{align}
	to better reflect the effect on the users. For minimal \textit{relative pool radius} $\tr=0$ ($r = 0$), users do not walk (door-to-door ride sharing). For $0<\tr<1$, the relative pool radius $\tr$ gives the percentage of the maximal trip length that users are required to walk.
	For maximal relative pool radius $\tr=1$ ($r = \lmax/2$) all users walk completely and no ride sharing takes place anymore. 
	
	As the pool radius increases, more users are not served and walk completely. The ratio $\omn$ of rejected stops, which is similar to the ratio of rejected users, follows directly from the fraction of trips with  lengths $\ell<2\,r$ by integrating the trip length distribution $\rho_{\textnormal{tl}}(\ell)$ only for rejected users, i.e.~from $0$ to $2\,r$, as
	
	\noindent \begin{align}
		\omn(\tr)&=\int\limits_0^{2\,r} \rho_{\textnormal{tl}}(\ell) \diff \ell=\left(\frac{2\,r}{\lmax}\right)^2
		=\tr^{\,2}\,.
		\label{eq:not_served_stops}
	\end{align}
	It only depends on
	the relative pool radius $\tr$. 
	
	When a request arrives, we assign it to one of $B$ buses and insert pickup and drop off stops (unless the user walks to other stops) into the current route of the bus. We determine the assignment and routing according to a simple algorithm that exclusively minimizes the bus route length, i.e.~the sum of the distances of all subsequent stops in its route. When a request appears, the algorithm calculates for each bus how to insert the origin and destination with minimal additional route length. For this purpose, it iterates over all currently planned stops in the bus route to check whether the user could be served indirectly via this planned stop (if $\tr > 0$) or, if not, how much an insertion of the new stop would increase the route length. In the end, the algorithm assigns the request to the bus with shortest route length after inserting the request. 
	If origin and destination are far from planned stops, the bus would pick up and deliver the user directly from and to their requested location. If there are planned stops near the origin or destination, the algorithm favors stop pooling to minimize the route length.
	
	The buses drive with velocity $\vv $ on the shortest path from stop to stop, serving all assigned users. Users walk to and from their pooled stops or their whole trip on the shortest path with velocity $\vp = \vv/10$.  
	For simplicity, we consider 
	buses with infinite capacity $c=\infty$ and 
	zero time to decelerate, park, serve users and accelerate again at each stop (zero stopping time). 
	This setting marks a lower bound for the efficiency of stop pooling because it can only save route length but no stopping time.
	
	For all simulations illustrated in the figures, we take a constant request rate $\lambda= 540$ and bus velocity $\vv=1$ ($\vp=0.1$) and vary the number of buses $B \in \{30, 35, \ldots, 60\}$ and the relative pool radius ${\tr \in \{0, 0.1, \ldots, 1\}}$ to analyze the influence of dynamic stop pooling on door-to-door ride sharing, modeled by $\tr=0$. 
	All other parameters are kept constant. In particular, we use exactly identical requests (request times, origins and destinations) in the different simulations, not only similar request distribution. In this way, we show how stop pooling can help to improve a certain service with given demand -- e.g.~in a given city.
	
	Clearly, stop pooling can only take place with stops of other users. Thus, users first have to share rides, before they can pool stops.
	The service is in the \textit{ride sharing regime}, i.e.~it has to bundle user trips, if more trip length is requested than the buses can travel per time. The \textit{load} 
	
	\noindent \begin{align}
		x =\frac{\lambda \, \lav}{ \vv \,B }\,
		\label{eq:load}
	\end{align}
	defined by
	Molkenthin et al.~\cite{Molkenthin.2020} characterizes the ride sharing regime by $x>1$. 
	Here, 
	$\lambda \, \lav$ is the average trip length requested per time and $\vv \,B$ the maximal distance all buses can travel together per time. The load is a lower bound for the average occupancy of the buses \cite{Molkenthin.2020}.
	As long as $x>1$, the buses are on average always occupied by at least one user and are almost never idle. 
	
	The higher the load $x$, the more user trips need to be bundled to serve all requests. However, high loads come along with high computation cost (of the assignment algorithm), high occupancy, and high user travel time and are unfeasible and unrealistic.
	To be well in the ride sharing regime without too high loads, we choose initial loads (door-to-door ride sharing) $x_0\in[3, 6]$ (compare parameters above). That means, three to six times more route length is requested than the buses can serve. In consequence, on average at least three to six users are in a bus per time step who can pool their stops. Due to detours and longer travel times, the actual occupancy is typically much larger, especially in settings with only few buses. 
	\subsection{Observables}
	
	We start our simulations with $B$ empty buses randomly distributed in the unit square and wait for some time $T_0 = 100$ 
	until the bus occupancy and the length of the planned routes have equilibrated. 
	We measure our observables in a fixed observation window $\Delta T = 100$, $t \in [100,200]$, in the steady state after equilibration. 
	In this window, approximately $P\approx \lambda\,\Delta T = 5\times10^4$ users are served. We consider only request with delivery in the observation window.  
	Because we simulate for such a long time and so many users, we observe well defined average values. The standard error of the mean for our observables is very small and thus negligible in the figures presented below.
	
	\subsubsection{Route length}
	\label{sec:l}
	
	The total route length $L$ is the sum of all bus route lengths. The route length $L_i$ of bus $i$ is the sum over all stop distances of the route of the bus.
	We normalize $L$ by the ideal total route length in individual mobility $L_\text{ind}$, the sum of all $P$ user trip lengths $\ell_j$. The rescaled observable\textit{ relative route length}
	
	\noindent \begin{align}
		\ltil&=\dfrac{L}{L_\text{ind}}=\dfrac{\sum\limits_{i=1}^B L_i}{\sum\limits_{j=1}^P \ell_j}=\frac{BvT}{P\langle \ell \rangle} \, \left(1-\langle p_\mathrm{idle}\rangle\right)\,.
		\label{eq:ltil}
	\end{align} 
	quantifies how much longer/shorter the buses drive to serve the users compared to each user going by car individually.
	Here, $\langle p_\mathrm{idle}\rangle$ is the probability for the buses to become idle.
	For $\ltil>1$, the buses would in total drive further than cars in individual mobility. 
	For $0<\ltil<1$, the service requires less bus route length to serve all users than individual mobility and is $1/\ltil$ times more efficient in route length. For $\ltil = 0$, no buses drive at all, the service does not serve anyone. This only occurs for $\tr=1$ when all users walk completely.
	
	Over a constant observation time $T$, the total route length by the bus fleet is directly proportional to the idle time of the buses. In particular, if buses are never idle due to sufficiently high load, $p_\mathrm{idle}\rightarrow 0$ for $x\rightarrow\infty$, the total route length $L \rightarrow B\,\vv\,T$ does not change with the relative pool radius $\tr$ or the load $x$. Similarly, the total sum of the user trip distance $L_\textnormal{ind}$ depends on the request rate $\lambda$ and the trip length distribution but not on $\tr$ such that the relative route length is independent of $\tr$ in the ride sharing regime.
	
	We take energy required and emissions caused to be proportional to the total route length driven, neglecting the influence of vehicle size or capacity compared to private vehicles. The relative route length $\ltil$ thus quantifies the energy consumption and emissions of a ride sharing system compared to ideal individual mobility. For $\ltil<1$, we thus consider the system to be ecologically more \textit{sustainable}.
	
	\subsubsection{Travel time}
	\label{sec:t}
	Usually, users pay for the reduced relative route length with longer travel times than in individual mobility. We measure the average of all $P$ user's travel time, which is the time between request and arrival at the destination. We normalize this average travel time by the ideal average travel time in individual mobility when all users are served immediately, without detour and with bus velocity $\vv$. This \textit{relative travel time} $\ttil$ reads
	
	\noindent \begin{align}
		\ttil=\dfrac{\langle t\rangle}{ \langle t_\textnormal{ind}\rangle}=\dfrac{\dfrac{1}{P} \sum\limits_{j=1}^P\left( t_{\mathrm{arrival},j}-t_{\mathrm{request},j}\right)}{
			\dfrac{\langle\ell\rangle}{\vv}}\,.
		\label{eq:ttravel}
	\end{align} 
	The  relative travel time measures how much slower users are compared to the ideal travel time. Because we measure a user related observable, we include all users into the  relative travel time. Rejected users simply contribute their walk time $t=\ell \ \vp$.
	The minimal possible  relative travel time in ideal individual mobility equals one. For $\ttil>1$, users are $\ttil$ times slower than in individual mobility.
	In the example study below we have $\tcp=\vv/\vp$ that measures the relative travel time when all users walk completely.

	\section{Results}
	\label{sec:results}
	
	\begin{figure*}[t]
		\includegraphics[width=\textwidth]{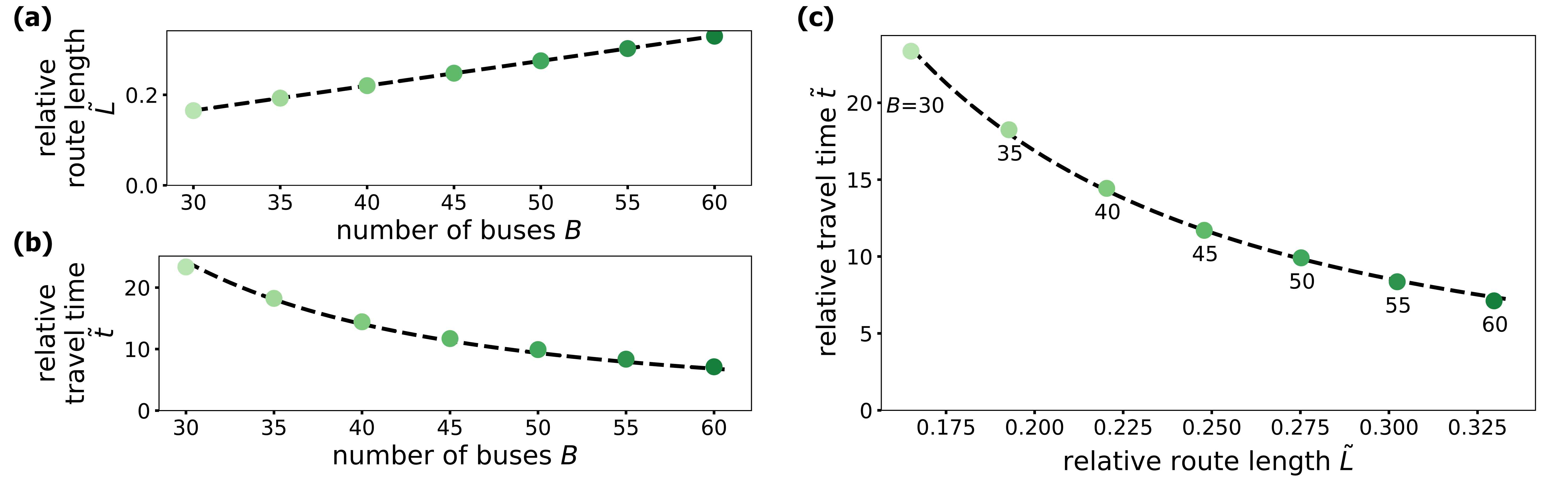}
		\caption{\textbf{Trade-off between relative route length and relative travel time in door-to-door ride sharing.}
			\textbf{(a)} The relative route length $\ltil$ increases approximately linearly with increasing number $B$ of buses  
			because $x \in [3,6]$ for constant $\lambda = 540$ are sufficiently high that buses are almost never idle ($\langle p_\mathrm{idle}\rangle \approx 0$, compare Eq.~\eqref{eq:ltil}).
			\textbf{(b)} At the same time, the  relative travel time $\ttil$ decreases. The fewer buses, the more route can be saved but the longer do users travel.
			\textbf{(c)} Relative travel time (from panel (b)) vs route length (from panel (a)) joined by similar $B$ encoded by small numbers beneath the curve.
			If reducing the  relative travel time, the relative route length increases in return and vise versa. It is hence impossible to improve both at the same time by just varying the number $B$ of buses in door-to-door ride sharing.
			Black dashed lines are guides to the eye.
		}
		\label{fig:ride_sharing_vs_B}    
	\end{figure*}
	\subsection{Door-to-door ride sharing}
	\label{sec:standardrs}

	First, we analyze how door-to-door ride sharing without stop pooling, ${\tr=0}$, with fixed request rate scales for different fleet sizes $B$.
	Relative travel time and relative route length scale oppositely: the relative route length increases with increasing number $B$ of buses (Fig.~\ref{fig:ride_sharing_vs_B}a); the relative travel time decreases with increasing $B$ (Fig.~\ref{fig:ride_sharing_vs_B}b). Joining these findings for similar $B$ shows that ride sharing services pay with increased relative travel time when reducing the relative route length by varying $B$ and vice versa (Fig.~\ref{fig:ride_sharing_vs_B}c). We thus identify a \textit{trade-off between relative route length and relative travel time} for door-to-door ride sharing. For given requests
	we cannot improve both at the same time (in analogy to \cite{AlonsoMora2018}).

	\subsection{Ride sharing with dynamic stop pooling}
	\label{sec:rswithsp}
	With dynamic stop pooling, users may walk to and from a close stop.
	For $\tr=0$, users do not walk (door-to-door ride sharing); for $\tr=1$, all users walk completely. Below, we explore the influence of any $\tr\in[0,1]$ on ride sharing in the model.
	
	\subsubsection{Fewer stops}
	\label{sec:fewerstops}
	
	The number of stops reduces in two ways: if users are served indirectly and walk to a nearby stop or if users are rejected and walk completely. 
	The second form of stop reduction is clearly undesirable for the users.
	Thus, the ratio of rejected users, which is the same as the ratio $\omn$ of rejected stops relative to the total number of stops, should be rather small, $\omn\ll1$. 
	
	The  ratio $\omn$ of rejected stops is proportional to the fraction of requests with destination in a circle with radius $2\,r$ around the origin, because these users are rejected and walk completely. With a uniform request distribution (see Sec.~\ref{sec:model}), $\omn$ grows quadratically in $r$ and is exactly equal to $\tr^2$ in terms of the normalized pool radius (see Sec.~\ref{sec:model}, Eq.~\eqref{eq:not_served_stops}).
	The ratio $\ome$ of served stops, which consists of the ratio $\omd$ of directly served stops and the ratio $\omi$ of indirectly served stops, thus decreases quadratically with $\tr$ as
	
	\noindent \begin{align}
		\ome(\tr)&=\omd(\tr)+\omi(\tr)=1-\omn(\tr)=1-\tr^{\,2}\,.
		\label{eq:effective_stops}
	\end{align}
	For minimal relative pool radius $\tr=0$ (door-to-door ride sharing), the buses serve all stops directly: ${\ome=\omd=1}$. For maximal pool radius $\tr=1$, the buses serve no stops ${\ome=\omd=0}$ and all users walk completely, ${\omn=1}$.
	Consequently, only small relative pool radii $\tr\ll 1$ are feasible so that most users are served. 
	
	\begin{figure*}[t]
		\includegraphics[width=\textwidth]{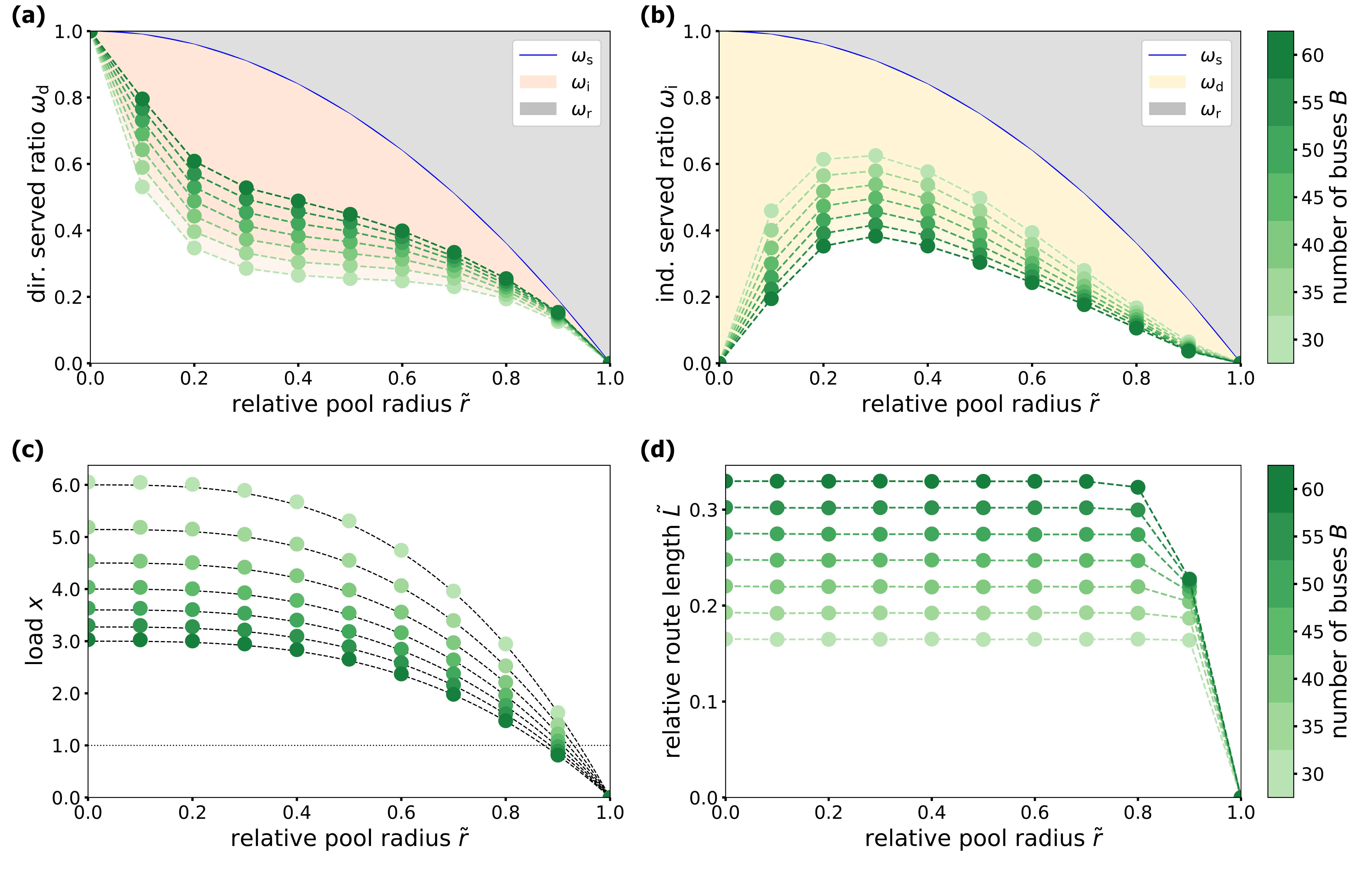}
		\caption{\textbf{Relative route length stays roughly constant although dynamic stop pooling saves stops.}
			\textbf{(a)}  
			The ratio $\omd$ of directly served stops decreases monotonically with $\tr$, faster than the served stop ratio $\ome$  
			(blue line, see Eq.~\eqref{eq:effective_stops}), which divides the saved stops into rejected $\omn$ (shaded grey) and indirectly served stop ratio $\omi$ (shaded orange).
			\textbf{(b)} 
			The indirectly served stop ratio $\omi$ first increases and then decreases again with increasing $\tr$. 
			Moreover, $\omi$ increases with decreasing $B$ (and constant $\lambda$) for ${0<\tr<1}$. That is, the more users share one bus, the more stops can be pooled. In general, stop pooling is only feasible for small pool radii $\tr\ll 1$ where most users are served ($\ome$, blue line) and the minority of users is rejected ($\omn$, shaded grey). 
			\textbf{(c)}~Rejections reduce the load $x$. For sufficiently small $\tr$, the load (black dashed lines according to Eq.~\eqref{eq:eload}) is high, $x>1$, and the system is in the ride sharing regime, such that all buses are busy at all times.
			For very high $\tr$, the load decreases to $x<1$ and almost no rides are served anymore. Buses have to wait for incoming requests.
			\textbf{(d)} Roughly constant relative route length $\ltil$ for small $\tr$
			due to busy buses for $x>1$ (cp.~Eq.\eqref{eq:ltil}). Only for sufficiently large $\tr$, the load falls below 1 (see panel (c)) and the route length decreases to zero when all users walk completely for $\tr=1$. 
		}
		\label{fig:Fig4}    
	\end{figure*}
	
	Simulations show that the ratio $\omd$ of directly served stops reduces with increasing relative pool radius faster than the ratio $\ome$ of served stops (Fig.~\ref{fig:Fig4}a). The remaining fraction $\omi$ of stops is served indirectly.
	This ratio $\omi$ of indirectly served stops quantifies the degree of actual stop pooling: how many stops are combined with others (instead of how many stops are rejected). For small pool radii, it increases and then decreases again with the relative pool radius when complete walking dominates (Fig.~\ref{fig:Fig4}b). First, more and more users walk to close stops with increasing relative pool radius. When the relative pool radius increases further, more and more of these users are rejected and walk their whole trip. Rejected stops replace indirectly served ones. 
	
	The potential of stop pooling increases with fewer buses. Since more users share a bus, the bus visits more stops that are on average closer together and can be pooled easier. Overall the more users share a bus, the higher is the potential of dynamic stop pooling.
	
	\subsubsection{Constant route length}
	\label{sec:roughlyconstL}
	
	\begin{figure*}[t]
		\includegraphics[width=\textwidth]{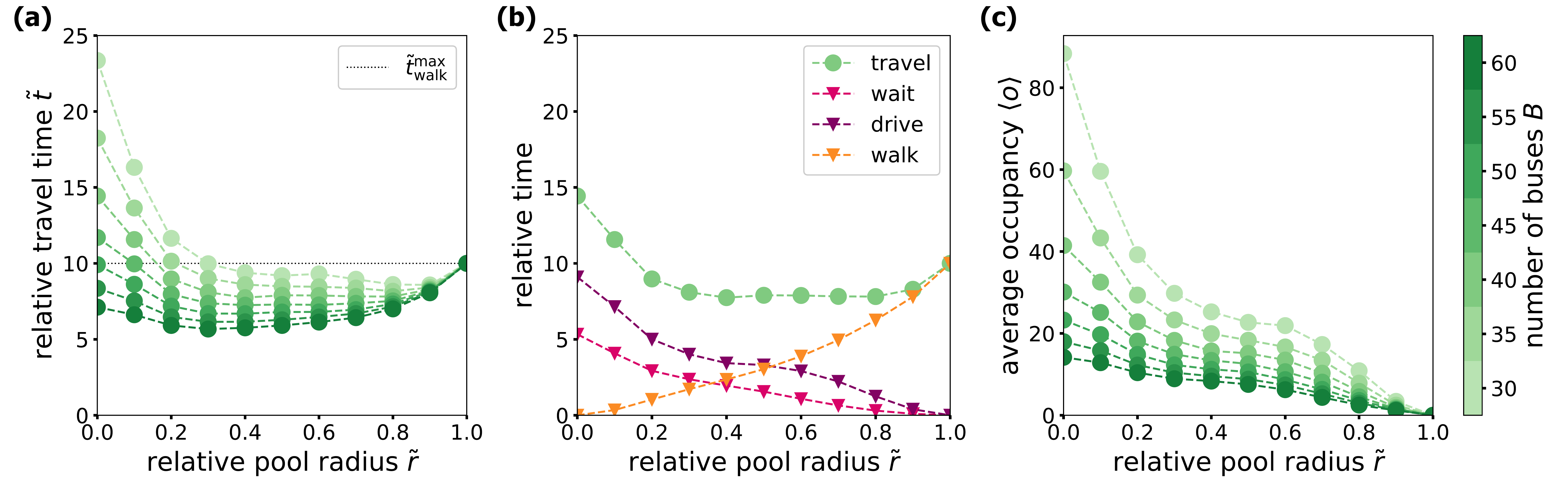}
		\vspace{-0.5cm}
		\caption{\textbf{Dynamic stop pooling reduces relative travel time and occupancy for sufficiently small $\tr$.} \textbf{(a)}~The relative travel time is minimal for some intermediate relative pool radius ($0<\tr<1$). This minimal relative travel time is lower than the relative travel time for door-to-door ride sharing ($\tr=0$) and lower than the relative travel time $\tcp$ (dashed line) when all users walk completely ($\tr=1$).
			In general, the relative travel time decreases with $B$ (as shown in Fig.~\ref{fig:ride_sharing_vs_B}).
			\textbf{(b)}~The relative travel time splits into drive, wait and walk time, shown here for $B=40$ (cp.~panel (a)). The wait and drive time decrease for increasing $\tr$. This effect is only partially explained by rejected users who walk completely and do not drive/wait. Additionally, buses avoid door-to-door detours, further reducing the drive and wait time of the remaining users. 
			Reduced drive and wait time overcompensate the increasing walk time for small enough $\tr$ such that the overall relative travel time decreases. For large $\tr$, the walk time dominates and the  relative travel time increases up to $\tcp$.
			\textbf{(c)}
			The average occupancy $\oav$ decreases with increasing $\tr$ and $B$. Dynamic stop pooling thus allows to use smaller buses than door-to-door ride sharing.
		}
		\vspace{-0.5cm}
		\label{fig:travel_time}    
	\end{figure*}
	
	The load $x$ measures how much trip length is requested compared to how far the buses can drive in total per time step (see Sec.~\ref{sec:model}, Eq.~\eqref{eq:load}). It is a lower bound for the average occupancy $\langle o \rangle$ of the buses \cite{Molkenthin.2020} - the average number of users per bus at any point in time.
	Because rejected users do not contribute to the load $x$, it depends on the relative pool radius as (derivation in Supplementary Material A)
	
	\noindent
	\begin{align}
		x(r)&=x_0 
		\left(1 -\tr^3\right)
		\label{eq:eload}
	\end{align}
	where $x_0$ denotes the load for door-to-door ride sharing with $\tr=0$. The load decreases with increasing relative pool radius 
	(see Fig.~\ref{fig:Fig4}c). Due to the high initial values $x_0\in[3,6]$, the load stays larger than one for most feasible pool radii. Consequently, the buses are typically occupied and thus remain busy almost all the time ($\langle p_\mathrm{idle}\rangle \approx 0$, compare Eq.~\ref{eq:ltil}). 
	Because they move with constant velocity, the buses drive the same route length in this time (observation window). Since the requests and their ideal total route length also stay the same, we measure a constant relative route length (Fig.~\ref{fig:Fig4}d). Only for (infeasibly) high relative pool radii close to one, the load falls below one. Buses become idle from time to time and wait for new requests without driving. The relative route length decreases until buses do not drive at all when all users walk at $\tr = 1$. 
	
	\subsubsection{Faster users}
	\label{sec:fasterusers}
	
	A constant relative route length despite saved stops might initially seem counter-intuitive.
	But the route length stays only constant from the point of view of the buses. Users see less of this route length, since they are faster and spend less time waiting for and driving in the buses (Fig.~\ref{fig:travel_time}a and b). 
	The relative travel time becomes minimal for some intermediate pool radius $0<\tr<1$ where neither all users are served from door to door ($\tr=0$) nor everyone walks ($\tr=1$). 
	
	Dynamic stop pooling can reduce the relative travel time 
	by making few users walk partially or completely and in turn reducing the drive and wait time. This reduction on average overcompensates the additional walk time for sufficiently small $\tr$ (Fig.~\ref{fig:travel_time}b).
	This comparison not only holds for the averages, but extends to the full travel time distributions as well (cp.~Supplementary Material~B~2).

	\subsubsection{Lower bus occupancy}
	\label{sec:lowero}
	
	Since users spend less time in the buses (see Fig.~\ref{fig:travel_time}a and b), the average occupancy $\oav$ of the buses reduces with dynamic stop pooling (Fig.~\ref{fig:travel_time}c). Fewer buses may serve the same requests with the same average occupancy that would have been impractically high for door-to-door ride sharing. For example $45$ buses require on average ${\oav=30}$ seats with $\tr$=0, but only ${\oav=18}$ with intermediate relative pool radius $\tr=0.2$, which could be served by a minibus.
	
	The magnitude of this effect increases the more users initially share a bus and goes beyond the pure rejections due to users walking completely (see Supplementary Material~C). When fewer buses serve the same requests, more users share the same bus such that dynamic stop pooling saves more stops, relative travel time and occupancy (Fig.~\ref{fig:Fig4}a and \ref{fig:travel_time}a and c).\\

	\begin{figure*}[t]
		\begin{minipage}[t]{0.49 \textwidth}
			\vspace{0cm}
			\centering
			\includegraphics[width=\textwidth]{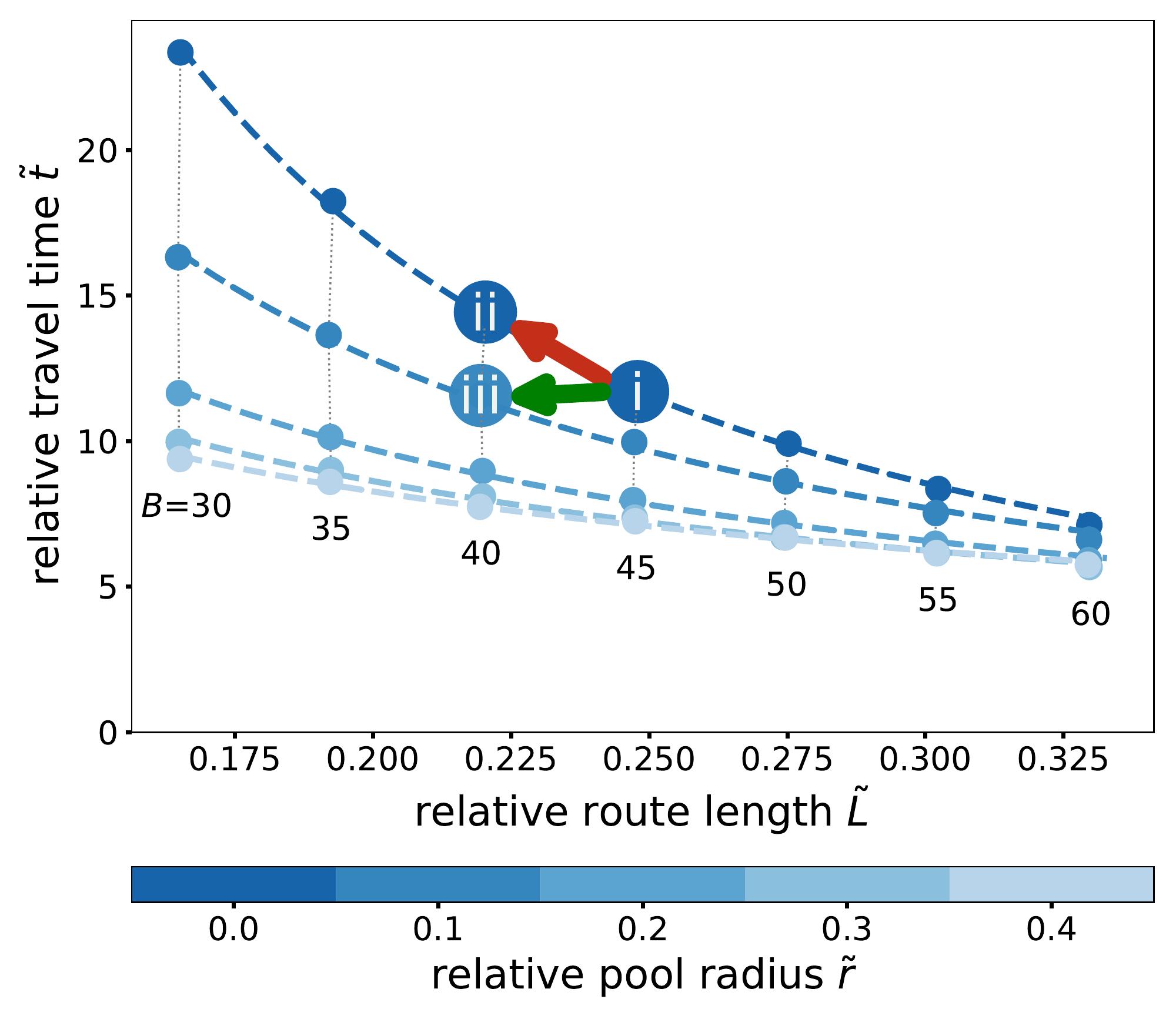}
		\end{minipage}
		\begin{minipage}[t]{0.49 \textwidth}
			\vspace{0.1cm}
			\small
			\begin{tabular}{lccc}
				\hline
				& \multicolumn{2}{c}{Ride sharing}&Stop pooling
				\\ 
				Scenario &         (i) & (ii) & (iii) \\
				\hline
				Relative pool radius $\tr$&0&0&0.1\\
				Number $B$ of buses &      45 &      40 &      40\\
				Load $x$&4.0361&4.5439&4.5370\\
				Route length $\ltil$ &  \cellcolor{red!15} 0.2479 &    0.2203 &   \cellcolor{green!15}  0.2195 \\
				Relative travel time $\ttil$ &       11.70 &      \cellcolor{red!15} 14.43 &     \cellcolor{green!15}   11.57 \\
				Average occupancy $\oav$ &       30.1 &      41.5 &       32.5 \\
				Stop ratio by type &&&\\
				\quad Directly served   $\omd$ &       1 &       1 &      0.64 \\\quad Indirectly served   $\omi$ &       0 &       0 &      0.35 \\\quad Rejected  $\omn$ &       0 &       0 &      0.01 \\
				User ratio by type &&&\\
				\quad Do not walk&1&1&0.56\\
				\quad Walk partially&0&0&0.43\\
				\quad Walk completely&0&0&0.01\\
				Walk length rel. to trip length &&&\\
				\qquad  by user type (in $\%$)&&&\\
				\quad Do not walk&$0$&$0$&$0.0 \pm \;\;0.0$ \\
				\quad Walk partially &-&-&$8.1 \pm \;\;7.8$ \\
				\quad Walk completely &-&-&$100.0 \pm \;\;0.0$ \\
				\hline
			\end{tabular} 
		\end{minipage}
		\caption{\textbf{Stop pooling breaks the ride sharing trade-off between relative route length and relative travel time.} 
			With fixed relative pool radius $\tr$ (shades of blue), the service shows the same trade-off between the relative travel time and the relative route length when varying $B$ (denoted by the small numbers beneath the data points), e.g.~for 
			(i)$\rightarrow$(ii) (cp.~Fig.~\ref{fig:ride_sharing_vs_B}). However, the increase in the  relative travel time is lower for higher relative pool radii. Increasing $\tr$ shifts the service to a lower relative travel time for sufficiently small $\tr$ (cp.~Fig.~\ref{fig:travel_time}). Together, increasing both $\tr$ and reducing $B$ results in decreased relative route length while keeping the relative travel time approximately constant, (i)$\rightarrow$(iii). Stop pooling thus breaks the ride sharing trade-off. Detailed data on the scenarios (i), (ii) and (iii) is given in the table on the right. Scenarios (i) and (ii) yield bad relative route length or relative travel time (red background), respectively. Scenario (iii) with dynamic stop pooling yields better results for both (green background). }
		\label{fig:2D}
	\end{figure*}
	
	In summary, with increasing (small enough) relative pool radius while keeping all other parameters constant (I) buses drive the same total route length because they are still busy all the time, (II) buses stop less often because more stops are pooled, thus (III) reducing the waiting time and detour for users and ultimately (IV) resulting in smaller average travel times for users \textit{despite} walking further.

	\subsection{Dynamic stop pooling breaks trade-off}
	\label{sec:tradeoffbreak}
	
	With fixed relative pool radius $\tr$, lowering $B$ reduces the relative route length (Fig.~\ref{fig:Fig4}d) but increases the  relative travel time (Fig.~\ref{fig:travel_time}a).
	The door-to-door ride sharing trade-off between relative route length and relative travel time when only varying $B$ persists with dynamic stop pooling for constant $\tr$ (Fig.~\ref{fig:2D}).
	Raising the relative pool radius $\tr$ with fixed $B$ decreases the relative travel time but keeps the relative route length roughly constant (as long as $\tr$ is feasibly small).
	However, in combination, it is possible to decrease the relative route length while keeping the relative travel time constant by reducing $B$ and raising $\tr$ simultaneously. 
	We no longer pay automatically with higher relative travel times for shorter relative route lengths. Dynamic stop pooling breaks the ride sharing trade-off between route length and travel time.
	
	This breaking of the trade-off is a qualitative novelty of dynamic stop pooling as opposed to static stop pooling or door-to-door ride sharing.
	Existing studies of static stop pooling (that focus on reduced route length and increases shareability) observed longer travel times \cite{Yao.2020, Lin2016, Ma.2013}. So far, only studies with dynamic stop pooling (including this article) have observed reduced travel times \cite{Fielbaum.2021,Mounesan.25.04.2021}.
	
	Instead of trading short bus route lengths for high user travel times, it is sufficient to let users walk a short part of their trips if both bus route and stop positions are flexible.
	
	To better understand this effect, consider the three scenarios illustrated in Fig.~\ref{fig:2D}. 
	In scenario (i), a door-to-door ride sharing service delivers the users with $45$ buses. 
	If the service provider decides to only use $40$ buses (scenario (ii)), the route length reduces by $11\%$, but users travel on average $23 \%$ longer due to the ride sharing trade-off. The provider risks losing users. 
	In scenario (iii), $40$ buses with dynamic stop pooling serve the users at the same speed as in scenario (i), if the users walk up to $10 \%$ of their maximal trip length. Both relative route length
	and relative travel time 
	are smaller than in scenarios (i) and (ii) (Fig.~\ref{fig:2D}). The provider saves $11\%$ route length, and $11\%$ of the buses without increasing the relative average travel time or requiring much larger buses. The same holds for the distributions of the user travel times (see Supplementary Material~B~2).
	
	In return, only one percent of the users walk completely. Around half of the users do not walk at all, and the remaining partially walking users walk on average $8\%$ of their trip. Averaged over all users, they walk only $5\%$ of their trip, thereby enabling a more sustainable ride sharing service with slightly faster users.
	
	\section{Discussion}
	\label{sec:discussion}
	
	The simple model introduced above captures fundamental spatio-temporal interaction of various agents of ride sharing systems with dynamic stop pooling. We systematically analyzed the collective system dynamics for varying pool radius and number of buses. Other parameters like the average bus velocity, the request rate and the average requested trip length are summarized in a system-wide load $x$ (Eq.~\eqref{eq:eload}), cf.~also \cite{Molkenthin.2020}. The results demonstrate that dynamic stop pooling may break the trade-off that is prevalent in door-to-door ride sharing systems between reducing relative bus route length and reducing relative user travel time (Fig.~\ref{fig:2D} and Sec.~\ref{sec:standardrs}).
	
	Dynamic stop pooling increases ride sharing efficiency by reducing the number of directly served stops (Sec.~\ref{sec:fewerstops}). It thereby avoids many small door-to-door detours and decreases the occupancy of the buses if some users walk a short distance to a dynamically determined stop, in contrast to static stop pooling where every user would have to walk to a prescribed stop. Dynamic stop pooling thereby decreases the relative travel time (Sec.~\ref{sec:fasterusers}) while keeping the relative route length constant (Sec.~\ref{sec:roughlyconstL}) -- a novel quality for ride sharing systems (Sec.~\ref{sec:tradeoffbreak}). As a consequence, upon increasing the maximum walk distance of dynamic stop pooling, a smaller number of buses may serve the same number of requests without longer travel times and without needing larger buses (Sec.~\ref{sec:tradeoffbreak}). Dynamic stop pooling may thus help to make ride sharing ecologically more sustainable by reducing the number of buses, resulting in lower energy consumption and emissions, without negatively impacting flexibility, service quality and travel times. 
	
	The general mechanism of breaking the ride sharing trade-off relies on the interplay of two general conditions: First, dynamic stop pooling is only possible in the ride sharing regime, $x>1$, where buses do not serve all users one after another, because (in a given time) the sum of the trip length of all users is larger than the distance that the buses are able to drive. Thus, multiple users share a bus, which allows to pool their stops. Since stops close to each other (in space and time) may be pooled, dynamic stop pooling becomes more efficient for higher loads $x$ (i.e.~for fewer buses or higher request rate) with many stops per bus, whereas the influence of dynamic stop pooling vanishes for small loads $x$ with few stops per bus. 
	
	Second, to neutralize the increase in the travel time with reducing $B$ (negative effect of the trade-off) the travel time needs to decrease enough with $\tr$. The travel time only decreases for small enough $\tr$ and up to some minimal value for each $B$. If reducing $B$ too much, even the maximal decrease of the travel time for optimal $\tr$ might not neutralize the increase in the travel time due to reduction of $B$ completely.
	For instance when reducing $B$ in the above example from 60 to 30, which would half the relative route length, however much increasing $\tr$ will yield a higher travel time (all $\ttil$ with $B=30$ are higher than $\ttil$ with $B=60$ and $\tr=0$, cp.~Fig.~\ref{fig:travel_time}). Dynamic stop pooling is unable to completely neutralize this high decrease in $B$ and only buffers it. But for small reduction in $B$, we can observe a shorter route length (due to $B$ reduction) without longer travel time when increasing $\tr$.
	The same effect could be observed if a shorter route length compensates an increasing travel time (cp.~\cite{Fielbaum.2021,Mounesan.25.04.2021}). This condition is typically fulfilled for small pool radii where the service avoids small door-to-door detours without rejecting a large fraction of users due to additional benefits to the remaining users (see Supplementary Material~C). High pool radii are not feasible since most users walk, increasing their travel time, and almost no served stops remain to be pooled. 
	
	These arguments hold under more general conditions than those studied in our simplified model. First of all, different assignment algorithms and different delay or capacity constraints may reduce the options to pool stops. Similarly, walking may not be possible for all users. While these aspects may limit the overall potential of dynamic stop pooling, it does not affect the qualitative  mechanisms described above. Moreover, additional aspects including substantial stopping times, deceleration and acceleration, and the influence of traffic density on lane-switching, overall vehicle velocity and stopping times may even increase the benefits of stop pooling in terms of added comfort and security. (For a more detailed discussion of the robustness, see Supplementary Material~D). 
	
	Our simple model setting may represent real world urban centers with high request densities at highly frequented locations although the continuous space strongly reduces the overlap of requested trips resulting in high relative travel times (see Supplementary Material Sec.~B 1).
	Equating the length and time scales in our model to typical conditions in Manhattan with a total area of $59,1 \,\mathrm{km}^2$ ($8\,\mathrm{km}$ per length unit) and average velocity $10\,\mathrm{km/h}$ \cite{NewYorkDOT2019}, the simulated request rate corresponds to $11.25$ requests per minute or less than $5\%$ of the typical taxi request rate in Manhattan (approximately $400000$ daily \cite{santi2014_shareabilityNetworks}). 
	Already for such a small fraction of requests and number of buses -- chosen for the sake of computational feasibility -- dynamic stop pooling may break the ride sharing trade-off. Furthermore, our results remain robust for larger request rates and numbers of buses. Indeed, sharing rides and pooling stops becomes even easier resulting in shorter relative travel times for comparable loads and similar relative savings from stop pooling (see Supplementary Material~D for supporting simulations).
	Moreover, we find the same qualitative result even if rejected users drive individually instead of walk (see also Supplementary Material~D). 
	
	Overall, we have identified the joint \textit{dynamic} interaction of walking, routing buses, and dynamically pooling stops as the core mechanism to break the ride sharing trade-off. 
	Better understanding the influence of dynamic stop pooling and the underlying mechanisms may thus help to enable simultaneously more sustainable and more flexible shared mobility.
	
	\normalsize
	
	\bibliographystyle{unsrt}
	\bibliographystyle{naturemag}
	\bibliography{bib_final}
	
	\subsection*{Acknowledgments}
	The authors thank Verena Krall,
	Felix Jung and all members of Chair for Network Dynamics for valuable discussions. This work was partially supported by the Volkswagen Foundation under grant no. 99 720 and the German Federal Ministry for Education and Research (BMBF). CL acknowledges support from the German Federal Environmental Foundation (Deutsche Bundesstiftung Umwelt DBU). The authors are grateful to the Centre for Information Services and High Performance Computing (ZIH) TU Dresden for providing facilities for high throughput calculations.
	\vspace{0.5cm}
	
	\subsection*{Code and data availability}
	The code and all data underlying the figures shown in this manuscript can be requested from the authors.

	\subsection*{Conflict of Interest}
	The authors declare no conflicts of interest.

	\onecolumngrid
	
	\null
	\newpage
	\renewcommand{\thefigure}{S\arabic{figure}}
	\renewcommand{\thetable}{S\arabic{table}}
	\renewcommand{\theequation}{S\arabic{equation}}
	
	\setcounter{figure}{0}
	\setcounter{table}{0}
	\setcounter{equation}{0}
	\setcounter{page}{1}
	\resetlinenumber

		\section*{Supplementary Material}
		\subsection{Load depends on relative pool radius}
		\label{sec:loadderivation}
		
		Rejected request do not use the ride sharing service and thus cannot contribute to the request rate $\lambda$ and the average trip length $\lav$. In consequence, both depend on the pool radius. The request rate simply reduces by the ratio of unserved users, which is similar to the ratio $\omn$ of unserved stops, as
		
		\noindent \begin{align}
			\lambda(\tr)&= \lambda_0  (1 - \omn(\tr)) = \lambda_0\left(1 -\tr^2\right)
			\,,
			\label{eq:e_request_rate}
		\end{align}
		
		\noindent
		where the index $0$ labels the respective quantity for $\tr=0$. 
		The average trip length for $\tr=0$ reads
		
		\noindent \begin{align}
			\lav_0 = \int\limits_{0}^{\lmax} \ell\,\rho_{\textnormal{tl}}(\ell)  \,\textnormal{d} \ell =\frac{2}{3}\lmax \overset{\lmax=1/2}{=}\frac{1}{3}\,,
			\label{eq:lav0}
		\end{align}
		where the integration runs over all users.
		For $\tr>0$, we have to exclude the trip length $\ell$ of the rejected users, which are those with $\ell<2\,r$. Thus, the integration runs only from $2\,r$ to $\lmax$. In addition, we have to re-normalize the trip length distribution due to the excluded users, which yields
		
		\noindent \begin{align}
			\lavr &= \int\limits_{2\,r}^{\lmax} \ell\,\dfrac{\rho_{\textnormal{tl}}(\ell) }{1 - \omn(r)} \,\textnormal{d} \ell = \int\limits_{2\,r}^{\lmax} \frac{2}{\lmax ^2}\dfrac{1}{1 - \omn(r)}  \ell^2 \,\textnormal{d} \ell= \dfrac{1}{1 - \omn(r)}\frac{2\ell^3}{3 \lmax ^2}\Biggr|_{2\,r}^{\lmax} 
			\nonumber \\
			&=\dfrac{1}{1 - \omn(r)}\frac{2}{3} \lmax \left(1-\left(\frac{2\,r}{\lmax}\right)^3\right)
			\nonumber\\	
			\lavtr&	\overset{\textnormal{Eq.~(2)}}{=\qquad}\dfrac{1}{1 - \omn(\tr)}\frac{2}{3}\lmax \left(1-\tr^3\right)\overset{\textnormal{Eq.~(3)}}{=}\dfrac{1}{1-\tr^2}\frac{2}{3}\lmax  \left(1-\tr^3\right)\nonumber\\
			&\overset{\textnormal{Eq.~(S2)}}{=\qquad} \dfrac{1-\tr^3}{1-\tr^2}\lav_0\,.
			\, \label{eq:cp_mean_trip_length}
		\end{align}
		The load $x$ (Eq.~(4)) contains both request rate and average trip length and thus depends on the relative pool radius as
		
		\noindent \begin{align}
			x(r)&=\dfrac{\lambda_0\left( 1-\tr^2\right)  \dfrac{1-\tr^3}{1-\tr^2}\lav_0}{\vv B}\nonumber
			\\&=\dfrac{\lambda_0\langle\ell\rangle_0}{\vv B} 	\left(1 - \tr^3\right)
			\nonumber\\&=x_0 
			\left(1 - \tr^3\right)\,
		\end{align}
		compare Eq.~2 in the main manuscript, where the number $B$ of buses and the bus velocity $\vv$ are constant with respect to $\tr$.
		
		\newpage
		\subsection{Travel time observations}
		\subsubsection{High relative travel times}
		For the fleet sizes considered in the main manuscript, our simple model yields high relative travel times (Fig.~5a). For small $B$, the relative travel time with ride sharing can even be larger than if all users walk their whole trip. Clearly, such high travel times are unrealistic and infeasible.
		
		There are two main reasons why users are this slow with few buses. First, we study the ride sharing dynamics in continuous space since it reveals the impact of stop pooling with continuously varying relative pool radius in isolation from additional effects from the network geometry or coarse graining. Continuous space constitutes a worst case scenario for sharing rides, since the stops of a user with probability zero lie on the direct route of any other user (or the planned routes of the buses). Thus, we observe worse detour and travel times than one could expect on a street network.
		
		Second, our assignment algorithm reduces the total route length that buses drive to optimize the throughput from the point of view of the service but without any regard for the user travel times. A more realistic algorithm would reject users (or users would reject the offers) when their predicted travel time is large or the request does not fit the currently planned bus routes. The current assignment scheme only rejects users whose direct walk distance is less than two pool radii.

		The basic setting considered in the main manuscript already demonstrates the qualitative effects of dynamic stop pooling. The reduction of user travel times will likely be even larger with a more optimized algorithm, stopping times, and stops constrained to a street network instead of continuous space, though the exact values will depend on the specific setting.

		\subsubsection{Travel time distributions}
		\label{sec:traveltimedist}
		
\begin{figure*}[ht]
			\includegraphics[height=6.45cm]{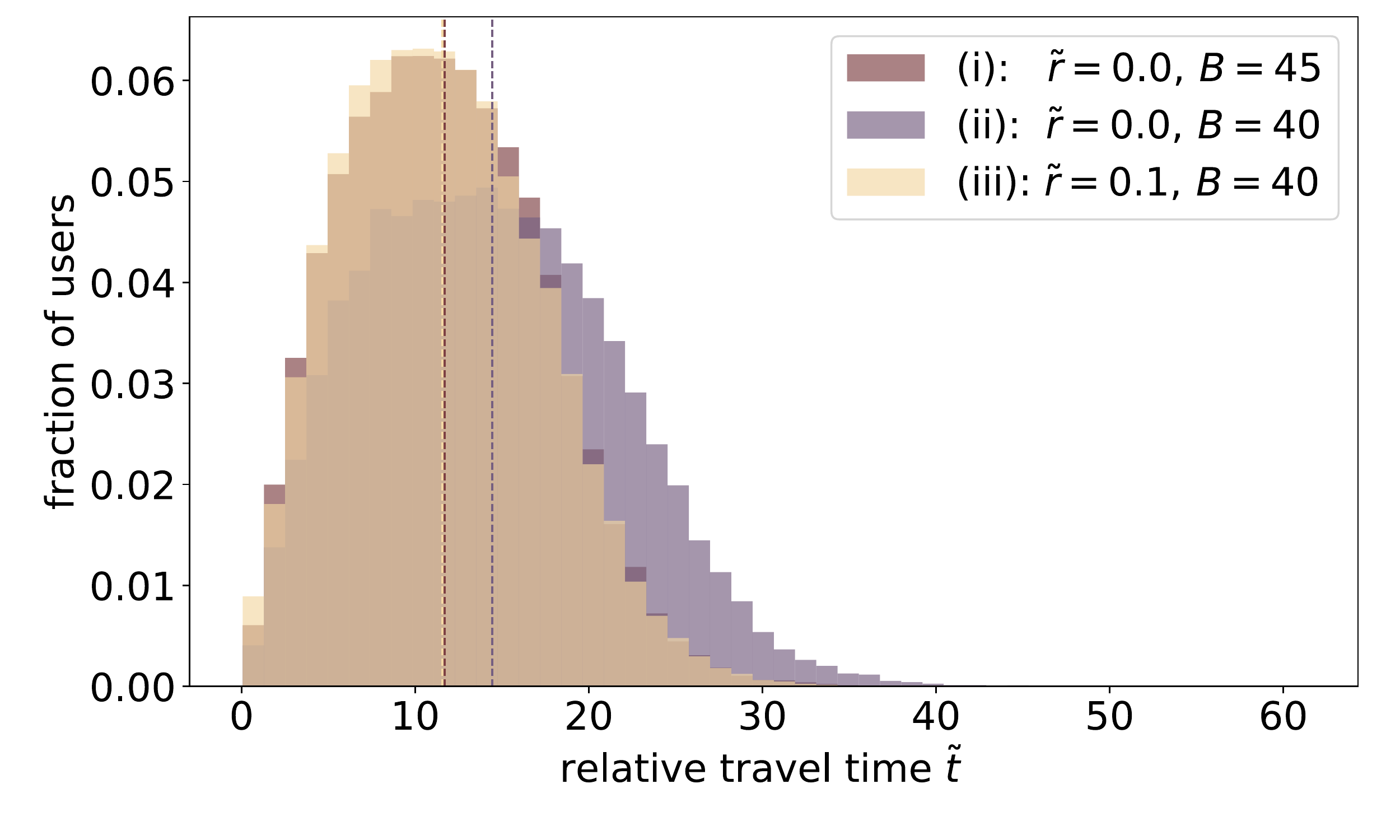}
			\caption{\textbf{Similar distribution for similar average travel time.}
				The fraction of users per travel time has a similar shape for all three scenarios discussed in Fig.~6 in the main manuscript and only differs in height and width. The respective average values (dashed lines) are thus a good measure for comparing the distributions.
			}
			\label{fig:travel_time_distribution}    
		\end{figure*}

		Fig.~5a and Fig.~6 in the main manuscript only display the average relative travel times. In figure \ref{fig:travel_time_distribution}, we show the distribution of the the individual travel time relative to the ideal average travel time of all users in individual mobility from the scenarios in  Fig.~6. The respective average values are marked as dashed lines. 
		Evidently, the shape of the distribution is similar for all scenarios. They only differ in height and width. Thus, the average travel time is a good measure to compare the distributions.
		The travel times are worst for scenario (ii) and best for scenario (iii) where they are slightly better than for scenario (i).
		
\begin{figure*}[ht]
			\includegraphics[height=6.45cm]{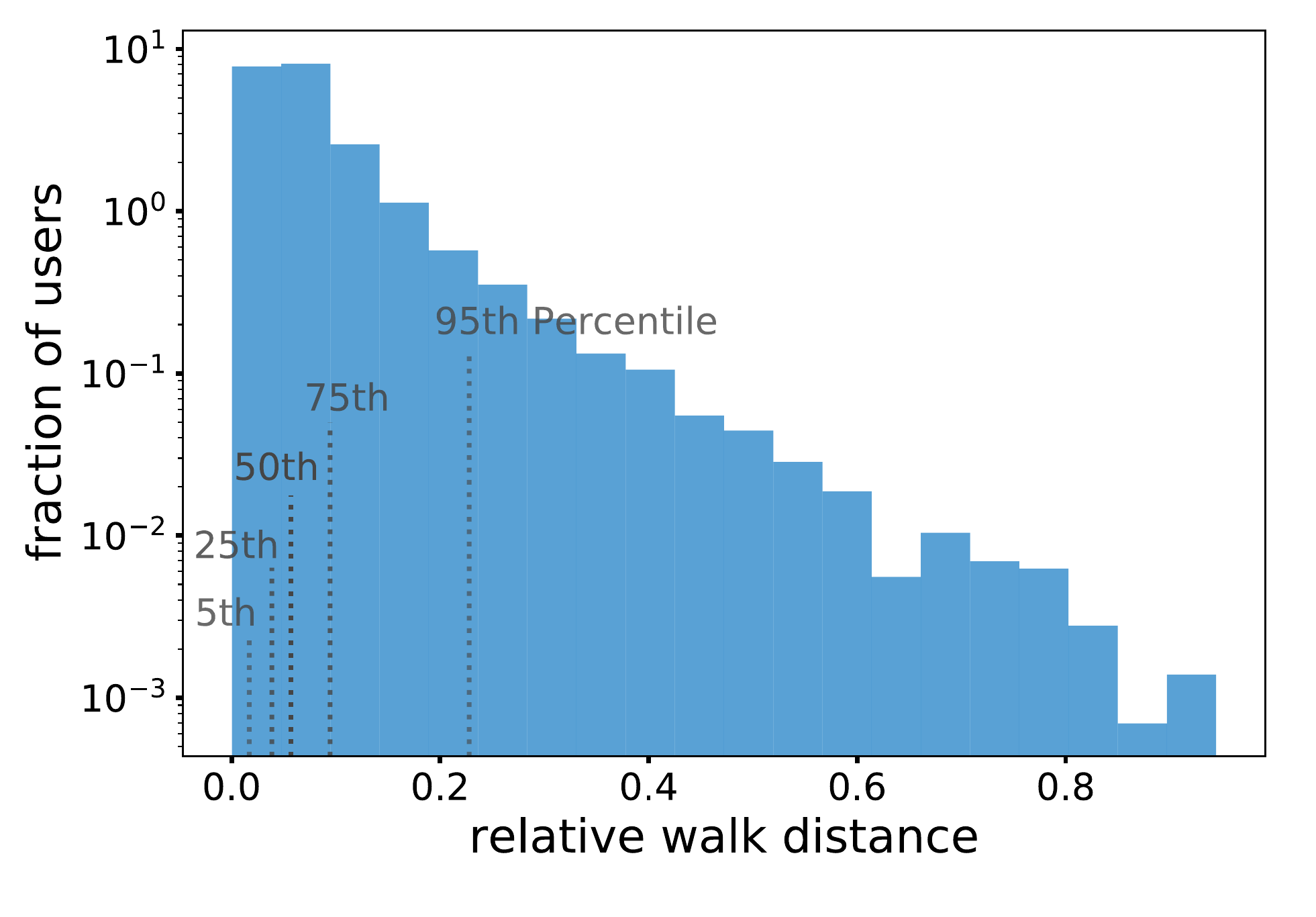}
			\caption{\textbf{Most users only walk a short part of their trip length.}
				Approximately exponential distribution of walk distance relative to the trip length for users in scenario (iii) for users that walk partially: Most users walk only a very short part of their trip: $95 \%$ of users walk less than  $21\%$ of their trip length (95th percentile); only $5\%$ walk further. Note the logarithmic scale on the $y$-axis.}
			\label{fig:walk_dist}    
		\end{figure*}

		In addition, we provide more detailed information on how far users walk in scenario (iii) from table in Fig.~6 ($B=40$,$\tr=0.1$) using the relative walk distance - the distance a user walks in total divided by the user's trip length. Users that do not walk have a relative walk distance of 0; rejected users have a relative walk distance of 1.
		For users that walk partially, we plot the distribution of the relative walk distance in Fig.~\ref{fig:walk_dist}. 
		We approximately find an exponential distribution. Most users only walk a short part of their trip length ($95\%$ walk less than $21\%$) and only very few users (often with very short trips) have to walk further.

		\newpage
		
		\subsection{Isolated effect of stop pooling}
		\label{sec:isolatedsp}
		
		\begin{figure*}[b]
			\includegraphics[width=\textwidth]{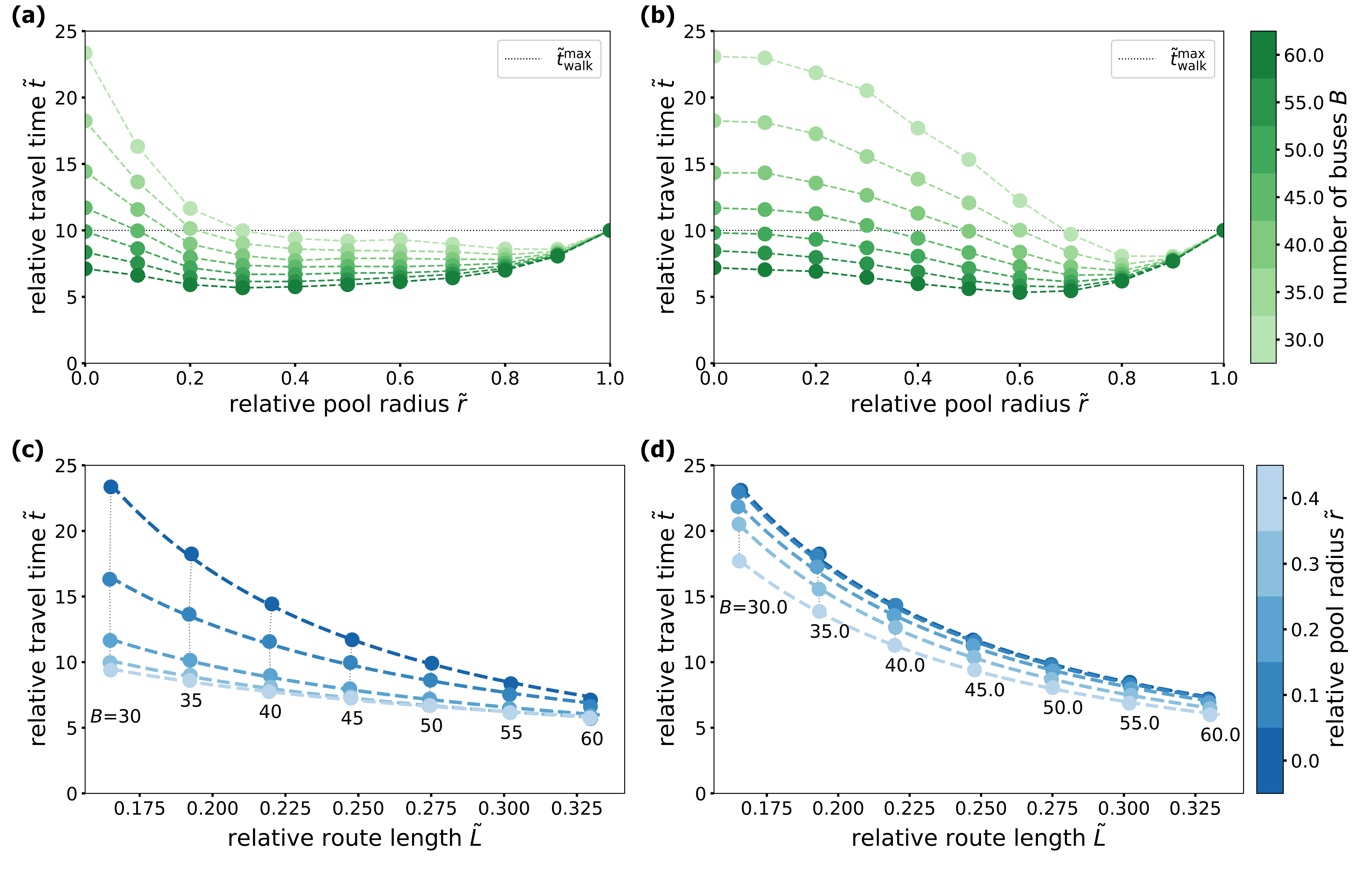}
			\vspace{-0.25cm}
			\caption{\textbf{Relative travel time reduces much more with dynamic stop pooling than only with rejections} for small pool radii.
				\textbf{(a)} The relative travel time reduces fast with increasing relative pool radius for ride sharing with dynamic stop pooling (equivalent to Fig.~5a). It becomes minimal for some intermediate relative pool radius, $0<\tr<1$. \textbf{(b)} Only with rejections, the  relative travel time does also decrease for small pool radii, but much slower than with dynamic stop pooling. The minimal relative travel time is higher and occurs for higher $\tr$ than with dynamic stop pooling. \textbf{(c)} Breaking of trade-off with dynamic stop pooling (equivalent to Fig.~6). For constant $\tr$, ride sharing services trade relative route length for relative travel time. The increase of the relative travel time curves decreases with $\tr$ such that the trade-off can be broken. \textbf{(d)} No clear breaking of trade-off only with rejections only. The increase in the relative travel time reduces slightly. The trade-off can only be broken for slight changes in $\ltil$ and with high increase in $\tr$.
			}
			\label{fig:isolated}    
		\end{figure*}
		
		In this section, we compare the results of ride sharing using dynamic stop pooling with simulations of door-to-door ride sharing with only rejections (users with $\ell<2\,r$ walk completely). In this way, we can isolate the effect from the pooled (indirectly served) stops.
		
		Because the load $x$ (Eq.~(8)) is only influenced by the rejections, it is similar with dynamic stop pooling and only with rejections and the relative route length stays roughly constant except for very high pool radii in both cases, since the buses are busy all the time for $x>1$ (Sec.~III B 2). 
		
		However, we find different influence on the relative travel time, which reduces faster and further with dynamic stop pooling than only with rejections (Fig.~\ref{fig:isolated}a,b). The relative travel time is much smaller with dynamic stop pooling for small relative pool radii - where the number of indirectly served stops is highest - whereas the reduction of the relative travel time is exclusively due to the reduced effective load only with rejections.
		
		In theory, a breaking of trade-off is possible in both cases because the increase of the relative travel time with reducing load decreases in both cases (Fig.~\ref{fig:isolated}). But, the relative travel time curves change only slightly with increasing $\tr$ only with rejections. To decrease the relative travel time with reduced relative route length, the relative pool radius has to increase by a lot. However, the area where the  relative travel time decreases with constant relative route length is limited (increase in relative travel time for higher relative pool radii). For this reason, the trade-off could only be broken in some boundary cases, where relative route length is only reduced slightly. 
		In contrast, the relative travel time curves flatten clearly with dynamic stop pooling, such that even a slight change in $\tr$  can  buffer the increase in the relative travel time due to a high reduction in relative route length by reducing $B$.
		
		Even if we correct Fig.~\ref{fig:isolated}c by excluding the effect of rejections from Fig.~\ref{fig:isolated}d, we still find the breaking of trade-off exclusively due to stop pooling effects. 
		For this reason, we expect that the breaking of trade-off will also be observed for models of dynamic stop pooling without rejections, for instance when the pool radius is adapted to the user's trip length.
		
		\newpage
		\subsection{Robustness}
		\label{sec:roustness}
		In this note, we discuss the simplifications mentioned in the main manuscript in more detail and explain why they do not alter the main result. We demonstrate that the trade-off break also holds for higher request rates and even if rejected users drive instead of walk.
		
		\subsubsection{Additional aspects}
		
		The model discussed in the main manuscript makes several simplifications to enable systematic analysis. Our main observation of breaking the trade-off is independent of these simplifications. Here, we discuss these individual aspects and their potential impact:
		\begin{enumerate}
			\item System size: The result is robust for higher request rates under constant load. We discuss this in more detail in the next subsection. Altering the system size by varying the size of the space or the velocities $\vv$ or $\vp$ only rescales time and/or space and does not affect the qualitative results since we only observe relative quantities (e.g. larger space with proportionally higher velocities rescales distances). Also altering the ratio of bus and user walk velocity  only scales the effect. Faster walking supports stop pooling and higher relative pool radii are feasible (and vice versa). 
			Of course, changing only space size or velocities will change the absolute values of travel time or route length.
			
			\item Assignment algorithm: The presented simple algorithm minimizes only the route length while serving all incoming requests. Additional constraints such as a maximum wait or travel time of the users may reduce the option to share rides. The ride sharing service then may not be able to fulfill all constraints and might reject requests even without stop pooling. Stop pooling may decrease the rate of rejections in these scenarios [19-21, 23, 24] instead of or in addition to reducing the overall travel times. We expect the qualitative influence to remain such that stop pooling increases service quality in terms of route length or fewer constraint-based rejections, without increasing the occupancy or route length. However, additional constraints in the assignment algorithm may reduce the options to pool stops such that the quantitative influence of dynamic stop pooling on ride sharing might be smaller. 
			
			\item Capacity of buses: In the model, buses have infinite capacity – which is of course far from applicable. But for $B\ge50$, buses are on average occupied by 14 to 23 users (Sec.~III B 4), which is a realistic capacity for ride sharing minibuses [18]. Moreover, the underlying continuum space constitutes a worst-case for the ride sharing service since no two requests are ever made from the same location, in contrast to a street network where possible origin and destination locations are restricted.
			Even restricting the capacity to low values between 3 and 9 still allows saving route length and maybe even travel time [23]. 
			Moreover, the reduction in average occupancy observed in the main manuscript may amplify the effect of dynamic stop pooling with limited capacity buses. For this reason, we assume that dynamic stop pooling breaks the trade-off even with finite capacity and this simplification does not change the results qualitatively.
			
			\item Stopping time: The model so far disregards that buses require time to decelerate, pick up or drop off users and accelerate again. In our model, dynamic stop pooling can thus only save time due to fewer detours. With a stopping time larger than zero, dynamic stop pooling would save additional time for each saved stop. The present model with zero stopping time thus analyses a minimal positive effect of dynamic stop pooling, which is already sufficient to demonstrate the breaking of  trade-off.
			
			\item Homogeneous pool radius: All users are required to walk up to a homogeneous pool radius independent of their trip length which is why the model rejects users with short trips. Below, we show that our results are robust regardless whether these rejected users walk their whole trip (slow but sustainable) or, contrary, drive by car (fast but unsustainable). Since users with very short trips may, in reality, not request a ride at all, the expected time saving of stop pooling with small pool radii may be even larger than in our simulations. Small pool radii would still enable stop pooling and avoid detours for users with longer trips but not result in any additional walk time for users with short trips. 
			
			\item Mean-field parameters: Since we study the steady state dynamics of the ride sharing service, our results do not necessarily extend to time-varying parameters such as rapidly fluctuating request rates during a day. The tested steady state setting may approximate fluctuations that are sufficiently slow through quasistatic adaption of request rates. Moreover, small variations of the request rate and distribution in time and space are implicitly included in the stochastic realization of the requests. 

		\end{enumerate}

		\newpage
		
		\subsubsection{Higher request rate}
		\label{sec:highlambda}
		
		In this section, we provide three scenarios for a higher request rate $\lambda = 1000$ (Tab.\ref{tab:2Dhighlambda}) in analogy to those of Fig.~6. We use $B\in[80,90]$ for comparable loads (double request rate requires double number of buses) and find the same mechanism for breaking the ride sharing trade-off.
		
		\begin{table}[ht]
			\caption{\textbf{Dynamic stop pooling can maintain travel time despite route savings}. In scenarios (i)' and (ii)', either the relative route length or the relative travel time are worse (red background) than in scenario (iii)'. With dynamic stop pooling, both relative route length and relative travel time are better (green background).
			}
			\label{tab:2Dhighlambda} 
			\vspace{-0.25cm}
			\small
			\begin{center}
				\begin{tabular}{lccc}
					\hline
					& \multicolumn{2}{c}{Ride sharing}&Stop pooling
					\\ 
					Scenario &         (i)' & (ii)' & (iii)' \\
					\hline
					Relative pool radius $\tr$&0&0&0.12\\
					Number $B$ of buses &      90 &      80 &      80\\
					Load $x$&3.6984&4.1605&4.1539\\
					Route length $\ltil$ &  \cellcolor{red!15} 0.24976 &    0.22205 &   \cellcolor{green!15}  0.22200 \\
					Relative travel time $\ttil$ &       7.67 &      \cellcolor{red!15} 9.46 &     \cellcolor{green!15}   7.54 \\
					Average occupancy $\oav$ &19.1&26.2&19.8\\
					Stop ratio by type &&&\\
					\quad Directly served   $\omd$ &       1 &       1 &      0.63 \\\quad Indirectly served   $\omi$ &       0 &       0 &      0.36 \\\quad Rejected  $\omn$ &       0 &       0 &      0.01 \\
					User ratio by type &&&\\
					\quad Do not walk&1&1&0.57\\
					\quad Walk partially&0&0&0.42\\
					\quad Walk completely&0&0&0.01\\
					Walk length rel. to trip length &&&\\
					\qquad  by user type (in $\%$)&&&\\
					\quad Do not walk&$0$&$0$&$0.0 \pm \;\;0.0$ \\
					\quad Walk partially &-&-&$9.9 \pm \;\;9.2$ \\
					\quad Walk completely &-&-&$100.0 \pm \;\;0.0$ \\
					\hline
				\end{tabular}  
			\end{center} 
			\vspace{-0.35cm}
		\end{table}
		
		In scenario (i), a door-to-door ride sharing service delivers the users with $90$ buses.
		If the service provider decides to only use $80$ buses (scenario (ii)), the route length reduces by $11\%$
		, but users travel 
		on average $23 \%$ longer. The provider risks to lose users.
		In scenario (iii), the 80 buses can serve the users at the same speed as in scenario (i) 
		with dynamic stop pooling, if the users walk up to $12 \%$ of their maximal trip length. Both relative route length 
		and  relative travel time 
		are smaller than in scenarios (i) and (ii) (Fig.~6). The provider saves $11\%$ route length without increasing the relative travel time and only loses a small fraction (about $1\%$) of users that are rejected and walk completely.
		In return, users walk on average only $5\%$ of their trip. Half of the users do not walk at all, one percent of the users walk completely and the remaining partially walking users walk on average $10\%$ of their trip.

		The relative savings are approximately similar to those found with lower request rate in the manuscript. However, the relative travel times in all three scenarios are significantly smaller than in the smaller fleet size scenarios presented in the main manuscript, although the load is 
		similar. In detail, scenario (i) ($x=4.0361$, $t=11.70$) has a lower load but higher relative travel time than scenario (ii)' ($x=4.1605$, $t=9.46$). While a higher request rate and proportionally large fleet size yields a comparable load, it also allows better ride sharing and stop pooling. 
		More vehicles and more requests make it more likely that two users requests similar trips or that a bus is already driving close to a requested trip. Thus, the shared trips of the requests in each bus and the pooled stops are more similar and the users experience less detour and walking distance.

		In summary, the effect of dynamic stop pooling is robust qualitatively for higher request rates but has lower quantitative impact on the relative travel times because they are already lower without stop pooling.
		
		\subsubsection{Rejected users drive instead of walk}
		\label{sec:cpdrive}
\begin{figure*}[ht!]	
			\begin{center}
				\includegraphics[width=\textwidth]{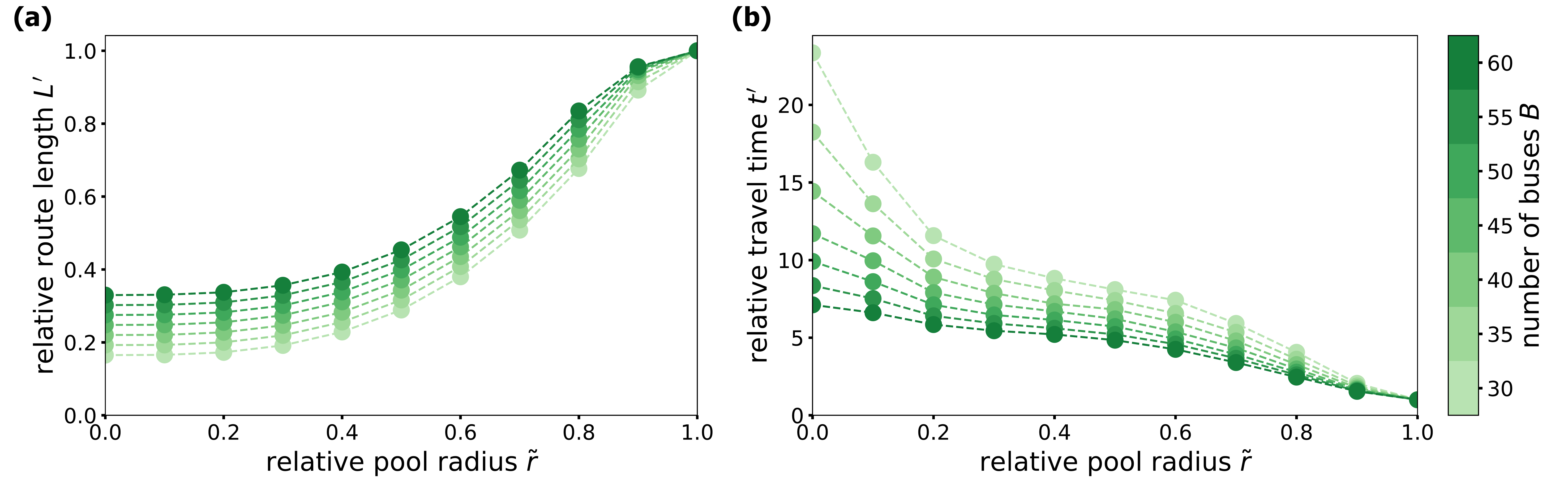}
			\end{center}
			\vspace{-0.25cm}
			\caption{\textbf{Only slight changes for small $\tr$ when treating rejections oppositely.}
				\textbf{(a)} The relative route length $L'$ if rejected users drive is larger than $\ltil$ for $\tr>0$. For $\tr=1$, it equals $L_\textnormal{ind}$. \textbf{(b)}  The relative travel time $t'$ if rejected users drive is smaller. For $\tr=1$, it decreases to $t_\textnormal{ind}=1/3$. 
			}
			\label{fig:cp_drive}    
		\end{figure*}

		In the model, we reject users with short trip lengths $\ell<2\,r$. Clearly, it is not desirable for users to walk completely if they request a transport service. However, this is a simple and computationally effective procedure to avoid that users walk further than their trip length. Here, we show that our results are robust regardless whether these rejected users walk their whole trip (slow but sustainable) or, contrary, drive by car (fast but unsustainable). Now, we let the users drive with $\vv$ without detour and wait time as if they would go by individual car. Thus, the single user travel time of each complete driving (cd) user can be calculated from their complete walking (cw) time times the velocity ratio
		
		\noindent \begin{align}
			t_{\textnormal{cd}} &= t_{\textnormal{cw}} \dfrac{ \vp}{ \vv}
		\end{align}
		We thus multiply all travel times of complete walk users by $\vp/\vv$ and add the additional route length / subtract the missing walk length.
		If we apply these changes for all $\omn$ complete walking users, we receive a higher route length, but a smaller travel time as
		
		\noindent \begin{align}
			L'-\ltil  &= \lcp \frac{P}{t_\textnormal{ind}}= \frac{2}{3}\lmax \tr^3 \frac{P}{t_\textnormal{ind}} 
			\label{eq:cp_drive_ldif}\\
			t'-\ttil&=\lcp  \left(\frac{1}{\vv}-\frac{1}{\vp}\right) \frac{1}{t_\textnormal{ind}}=\frac{2}{3}\lmax \tr^3 \left(\frac{1}{\vv}-\frac{1}{\vp}\right) \frac{1}{t_\textnormal{ind}}
			\label{eq:cp_drive_tdif}
		\end{align}
		where $L',t'$ denote the observables with complete drive instead of complete walk $\ltil,\ttil$ (previously presented). These outcomes are illustrated in Fig.~\ref{fig:cp_drive}.
		
		We find that by contrast to complete walk, the relative route length $L'$ does not decrease to $0$ but to $L_\textnormal{ind}$ for $\tr=1$, since each rejected user contributes his trip length. At the same time, the relative travel time $t'$ reduces even further to $t_\textnormal{ind}$ for $\tr=1$, since rejected users travel with higher velocity. The qualitative behavior for small $\tr$ remains unchanged. The relative route length is approximately constant since the completely driving user do not contribute much. The travel time decreases quickly due to stop pooling. Thus, the main mechanism underlying the trade-off break remains unchanged. However, large pool radii become unsustainable (identical to individual motorized mobility), instead of slow for completely walking users.
		
		What does this mean for the scenarios from Fig.~6? Table~\ref{tab:2D_cp_drive} summarizes the results for compete drive instead of complete walk. Since $\ltil$ and $\ttil$ only change slightly for small $\tr$, there is almost no effect of complete drive on the scenarios. Even if we consider the increase in $L'$ it can be overcompensated by the decrease in $t'$ such that we can still find a scenario with better $L'$ and not worse $t'$ for all door-to-door ride sharing scenarios.
		This shows that the previously presented results are robust to different handling of users with $\ell<2\,r$, because both extremes yield in principle similar outcomes (breaking of trade-off) if users drive instead of walking completely.
		
		\begin{table}[ht]
			\caption{Scenarios for completely driving rejected users (in analogy to Fig.~6) with the same overall result: In scenarios (i) and (ii) either relative route length or relative travel time are worse (red background) than in scenario (iii) (green background). For Scenarios (i) and (ii) with $\tr=0$ nothing changes since no users walk or drive completely. But for scenario (iii) with $\tr=0.1$, the relative route length is slightly worse if users drive with their own car than if users walk completely (but still much better than for scenario (i)); the relative travel time is slightly better. 
			}
			\label{tab:2D_cp_drive}    
			\begin{center}
				\begin{tabular}{lccc}
					\hline
					& \multicolumn{2}{c}{Ride sharing}&Stop pooling
					\\Scenario&         (i) &         (ii)&          (iii) \\
					\hline
					Relative pool radius $\tr$&0&0&0.1\\
					Number $B$ of buses &      45 &      40 &      40\\
					Relative route length $L'$ &  \cellcolor{red!15} 0.2479 &    0.2203 &   \cellcolor{green!15}  0.2205 \\
					Relative travel time $t'$ &       11.70 &      \cellcolor{red!15} 14.43 &     \cellcolor{green!15}   11.56 \\
					\hline
				\end{tabular}
			\end{center}
		\end{table}

\end{document}